\documentclass[aps,prb,twocolumn,groupedaddress]{revtex4}
\usepackage{graphicx}
\begin{document}


\title{The pressure dependence of many-body interactions in the
organic superconductor $\kappa$-(BEDT-TTF)$_{2}$Cu(SCN)$_{2}$: A comparison
of high pressure infrared reflectivity and Raman scattering
experiments.}


\author{R.~D.~McDonald$^{1,2,}$} \email[]{rmcd@lanl.gov}
\thanks{Ross McDonald would like to thank Drs. Neil Harrison, Charles Mielke and Albert Miglori for stimulating and fruitful discussions.} \author{A.-K.~Klehe$^{2}$ and J.~Singleton$^{1,2}$ and W.~Hayes$^{2}$.}
\affiliation{$^{1}$NHMFL, Los Alamos National Laboratory, MS-E536, NM~87545. USA. \\
$^{2}$Clarendon Laboratory, Department of Physics, Parks Road, Oxford, OX1~3PU, UK.} 




\date{June 13, 2002}

\begin{abstract}
We determine the pressure dependence of the electron-phonon
coupling in $\kappa$-(BEDT-TTF)$_{2}$Cu(SCN)$_{2}$ by comparison
of high pressure Raman scattering and high pressure infrared (IR)
reflectivity measurements. The Raman active molecular vibrations
of the BEDT-TTF dimers stiffen by 0.1-1~$\%$GPa$^{-1}$. In
contrast, the corresponding modes in the IR spectrum are observed
at lower frequency, with a pressure dependence of
0.5-5.5~$\%$GPa$^{-1}$, due to the influence of the
electron-phonon interaction. Both dimer charge-oscillation and
phase-phonon models are employed to extract the pressure
dependence of the electron molecular-vibration coupling for these
modes.
Analysis of our data suggests that the reduction of
electron-phonon coupling under pressure does not account for the
previously observed suppression of superconductivity under
pressure and that electron-electron interactions may contribute
significantly to the pairing mechanism.

\end{abstract}



\maketitle

\renewcommand\topfraction{1}
\renewcommand\textfraction{0.00}

\begin{sloppypar}

\section{Introduction}
$\kappa$-(BEDT-TTF)$_{2}$Cu(SCN)$_{2}$ is one of the best
characterized organic superconductors~\cite{Singleton_RPP2000}. It
is a highly anisotropic material with a quasi-two dimensional band
structure, whose Fermi-surface topology has been determined by
magnetotransport
experiments~\cite{Singleton_RPP2000,Caulfield_JPCM1994}. At
ambient pressure $\kappa$-(BEDT-TTF)$_{2}$Cu(SCN)$_{2}$ is a
superconductor with a transition temperature of $T_{\rm c}\simeq
10.4$~K. $T_{\rm c}$ decreases upon the application of pressure
until, at pressures $P$ exceeding 0.5~GPa, superconductivity is
fully suppressed~\cite{Caulfield_JPCM1994,Murata_SM1989,
harrison4}. The quasiparticle effective mass, $m^{*}$, derived
from magnetic quantum oscillation measurements, decreases linearly
with pressure up to 0.5~GPa; above this pressure the magnitude of
d$m^{*}$/d$P$ is strongly reduced~\cite{Caulfield_JPCM1994}. The
effective mass measured in this fashion includes contributions
from both electron-phonon and electron-electron interactions
~\cite{Legget_AoP1968}.

In contrast, the optical mass, $m_{\rm opt}$, extracted from a sum
over the optical conductivity, decreases approximately linearly
throughout the above pressure range~\cite{Klehe_JPCM2000}. This
mass is thought to be dominated by intraband electronic processes,
reflecting the band mass without renormalization by
electron-electron and electron-phonon
interactions~\cite{Legget_AoP1968,endnote1}. The coincidence of a
`kink' in the pressure dependence of $m^{*}$ with the pressure
above which superconductivity is suppressed and the absence of a
`kink' in the pressure dependence of $m_{\rm opt}$, suggests that
the interactions parameterized by $m^{*}$ maybe associated with
the superconductivity.

\begin{figure}[ht]
\includegraphics[width=8.5cm]{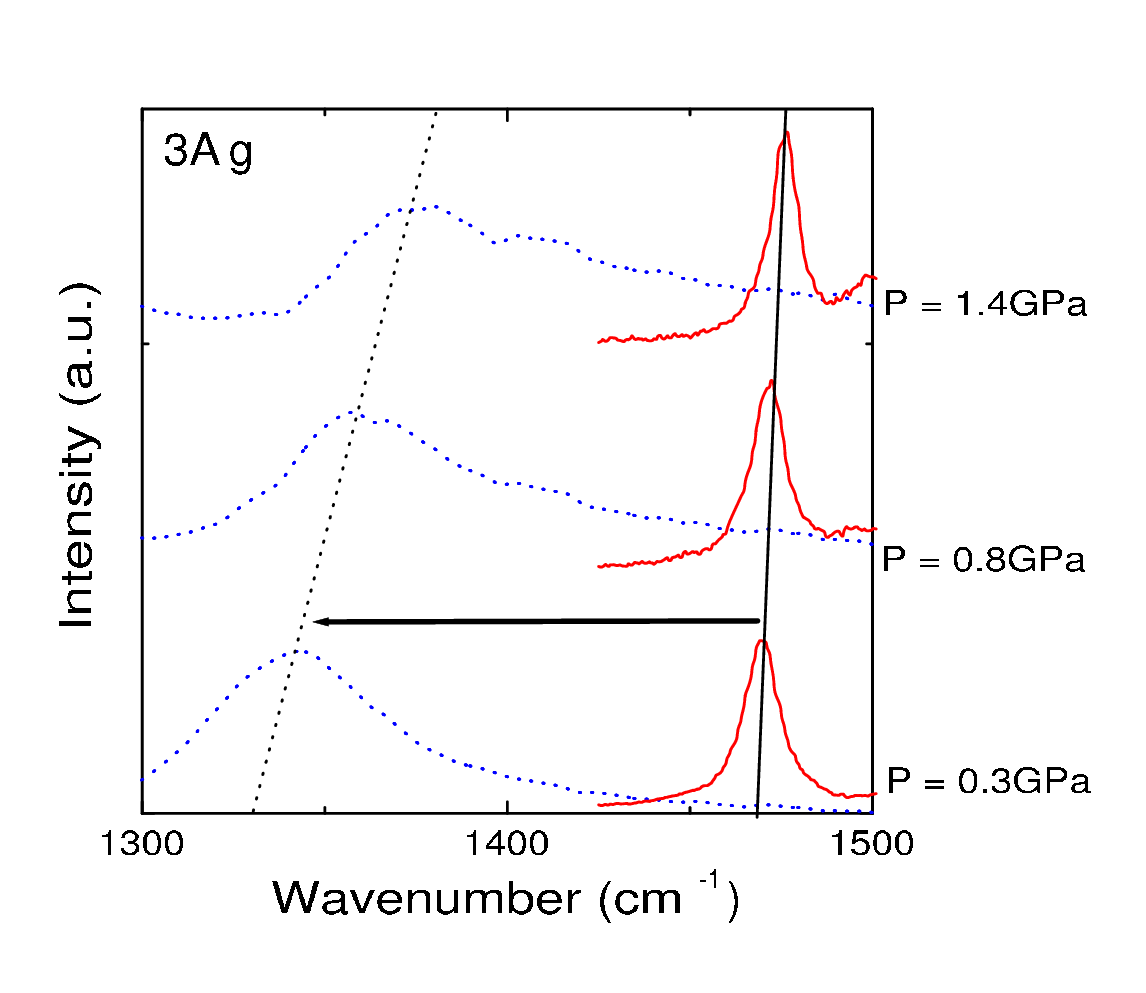}
\centering \caption{The spectral region containing the 3A$_{g}$
molecular vibration, Raman spectra (solid lines) and Infrared
spectra with polarization parallel to the crystallographic {\it
b}-axis (dotted lines), for a selection of pressures at room
temperature. The spectra are offset for clarity and the vertical
lines are a guide to the eye. Note the substantial softening
effect of the increased electron-phonon interaction on the IR
spectra with respect to the Raman spectrum.} \label{IRRfig1}
\end{figure}

The key question to address is the effect of these interactions,
{\it i.e.} what is the dominant pairing mechanism for
superconductivity in this material? In this paper we compare
infrared (IR) \cite{Klehe_JPCM2000,IR plane} and Raman scattering
\cite{McDonald_JPCM2001} measurements under pressure to determine
the role of the electron-phonon interaction. This is possible
because the IR measurement probes the molecular vibrations dressed
by the electron-phonon interaction
\cite{Kornelsen_SSC1989,Sugano_PRB1989}, whereas non-resonant
Raman measurements probe the bare mode frequencies
\cite{Sugano_PRB1989,Rice_SSC1979}. The primary effect of the
electron-phonon interaction is to soften the IR modes with respect
to the Raman modes, as illustrated in Figure~\ref{IRRfig1}.

There are two approaches to modelling the electron-molecular
vibration coupling interaction and associated mode softening in
organic charge transfer systems, the ``phase-phonon" theory and
the ``dimer charge-oscillation" model. The phase-phonon theory
approaches the problem from an electron band approximation where
the charge carriers are naturally delocalized. The alternative
dimer charge-oscillation theory is formulated in real space such
that the electronic behaviour is implicitly localized. In reality
the electronic properties of
$\kappa$-(BEDT-TTF)$_{2}$Cu(SCN)$_{2}$ are somewhere between the
extremes of a localized and itinerant
system~\cite{Singleton_RPP2000}. For this reason we apply both the
dimer charge-oscillation and phase-phonon models to analyze the
data.

This paper is organized as follows. In Section~\ref{modesS} we
describe the vibrational spectra of
$\kappa$-(BEDT-TTF)$_{2}$Cu(SCN)$_{2}$ as observed by both
non-resonant Raman scattering and polarized IR reflectivity.
Section~\ref{ssAnalysis} describes the methods of analysis,
specifically how the pressure dependence of the electron-phonon
coupling is extracted. The first subsection of the Discussion,
Section~\ref{M*}, contains a comparison of the quasiparticle
masses that result from the different methods of analysis.
Subsequently, in Section~\ref{TcSS}, in light of the pressure
dependence of the electron-phonon coupling, the implications for
the superconducting pairing mechanism are discussed. Conclusions
are summarized in Section~\ref{conc}.

\section{\label{modesS}Description of the observed vibrational modes.}
\begin{figure}[ht]
\includegraphics[width=8cm]{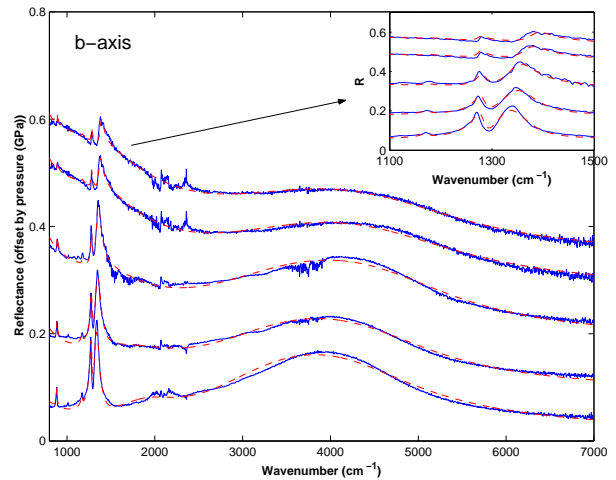}
\centering \caption{The room temperature diamond/sample IR
reflectance offset by pressure for polarization parallel to the
{\it b}-axis. Measurement (solid lines) and phase-phonon fit
(dashed lines). The inset magnifies the spectral region containing
the C=C mode.} \label{PPfitb}
\end{figure}

\begin{figure}[ht]
\includegraphics[width=8cm]{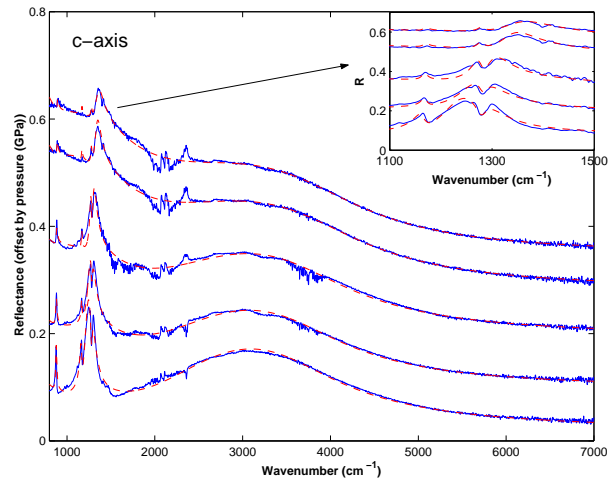}
\centering \caption{The room temperature diamond/sample IR
reflectance offset by pressure for polarization parallel to the
{\it c}-axis. Measurement (solid lines) and phase-phonon fit
(dashed lines). The inset magnifies the spectral region containing
the C=C mode.} \label{PPfitc}
\end{figure}

Four modes are observed in both the high-pressure room-temperature
Raman and IR spectra (see Figures~\ref{PPfitb} and \ref{PPfitc}
and Table~\ref{wtable}). They are labeled with subscripts
indicating the atoms/bond predominantly involved in the vibration
\cite{Eldridge_SAA1995}. In order of increasing frequency they are
the C-S mode originating from the BEDT-TTF 60B$_{3g}$ asymmetric
vibration, the central C=C mode originating from the BEDT-TTF
3A$_{g}$ symmetric vibration and two Cu(SCN)$_{2}$ anion modes.
Determining the pressure dependence of the central C=C mode in the
IR spectrum is not straightforward due to the overlap of several
modes with varying pressure dependences. In a recent publication
\cite{McDonald_SM2001}, IR spectra from the deuterated salt have
been used to model this Fermi resonance and extract the linear
pressure dependence of the overlapping modes (the central C=C mode
and two C-H modes). The IR and Raman frequencies and the first
order pressure dependence of all modes discussed here are given in
Table~\ref{wtable}.

The Cu(SCN)$_{2}$ anion modes do not exhibit any evidence of
electron-phonon coupling, {\it i.e.} the difference between the
infrared and Raman frequencies for the anion modes is accounted
for by purely vibrational coupling~\cite{McDonald_JPCM2001}. The
zone center frequency separation, $\Delta\omega_{\rm s}$, between
the symmetric (Raman active) and antisymmetric (infrared active)
combinations of the molecular vibrations is determined by the
frequency of the optical branch of the lattice modes
\cite{Zallen_PRB1974,McDonald_JPCM2001}. The larger observed
$\Delta\omega_{\rm s}$ for the anion mode in the {\it b}-axis
response, for which the lattice mode is stiffer (see
Table~\ref{wtable}), is a further confirmation of the lattice mode
assignment given in \cite{McDonald_JPCM2001}.
\begin{table*}[htb]
\begin{center}
\begin{tabular}{cccccccc} \hline \hline
 Mode & Assignment & \multicolumn{2}{c}{Raman} & \multicolumn{2}{c}{Infrared {\it b}-axis} & \multicolumn{2}{c}{Infrared {\it c}-axis}\\
 &$[$\onlinecite{Eldridge_SAA1995}$]$& cm$^{-1}$&+(\%GPa$^{-1}$) & cm$^{-1}$&+(\%GPa$^{-1}$) & cm$^{-1}$&+(\%GPa$^{-1}$)\\
  \hline\hline
 $\omega_{\rm CT}$ &$\cdots\cdots$ & $\cdots\cdots$ & $\cdots\cdots$ & 2910 &+ 4.0 & 2390 &+ 4.0\\
   $\omega_{\rm CS}$ & B$_{3g}$& 886.2 &+ 0.85& 883.5 &+ 0.71 & 873.6 &+ 1.0\\
   $\omega_{\rm CC}$ & A$_g$&  1467.7 &+ 0.4& 1290 &+ 2.5 & 1210 &+ 5.5\\
    $\omega_{\rm CH_{1}}$ & B$_{3u}$ & $\cdots\cdots$ & $\cdots\cdots$ & 1181 &+ 0.5 & 1177 &+ 0.5\\
     $\omega_{\rm CH_{2}}$ & B$_{1u}$ &$\cdots\cdots$ & $\cdots\cdots$& 1290 &+ 0.5 & 1281 &+ 0.5\\
     $\omega_{\rm anion1}$ & CN &2064.6 &+ 0.1 & 2067.4 &+ 0.1 & 2065.6 &+ 0.15\\
     $\omega_{\rm anion2}$ & CN &2106.3 &+ 0.2 & $\cdots\cdots$& $\cdots\cdots$ & 2109.3 &+ 0.2\\
      \hline
\end{tabular}
\caption{Raman and IR frequencies and pressure shifts taken from
\cite{Klehe_JPCM2000,McDonald_JPCM2001,McDonald_SM2001}.} \label{wtable}
\end{center}
\end{table*}

\section{Analysis.\label{ssAnalysis}}
Electron-molecular vibration coupling can be characterized by a
set of linear coupling constants, $g_{\alpha}$, which parameterize
the interaction between the $\alpha^{\rm th}$ normal mode, of
frequency $\omega_{\alpha}$, and the molecular orbital in which
the radical electron or hole resides \cite{Yartsev_IJMPB1998}. For
a donor molecule $g_{\alpha}$ is the rate of change of the energy
of the highest occupied molecular orbital with respect to the
normal mode coordinate, $Q_{\alpha}$, \cite{Yartsev_IJMPB1998}
\begin{equation}
g_{\alpha} = \frac{\partial {\rm E}}{\partial Q_{\alpha}}.
\end{equation}
To obtain these coupling constants experimentally, an appropriate
model has to be applied to the data. Two alternative approaches
exist in the literature: the ``phase-phonon" theory and the
``dimer charge-oscillation" model. Both theories were initially
developed by Rice \cite{Rice_PRL1976,Rice_SSC1979} to model the
infrared spectrum of semiconducting organic charge-transfer
systems. However, they have also been applied to a range of
reduced dimensionality organic charge transfer salts, including
metallic systems
\cite{Yartsev_IJMPB1998,Yartsev_JPCM1990,Rice_SSC1977,Rice_PRB1979,Sugano_PRB1989,Bozio_JCP1982,Kornelsen_PRB1991}.

The phase-phonon theory approaches the problem from an electron
band approximation where the charge carriers are naturally
delocalized. A potential disadvantage of this approach is that it
is a one-electron theory, {\it i.e.} all electron-electron
correlation effects have to be included within an effective mass
approximation \cite{Yartsev_IJMPB1998}.

The alternative dimer charge-oscillation theory is formulated in
real space but is limited to a finite number of molecules. The
infrared response of the system is calculated from a superposition
of the isolated units with no significant charge transfer between
the units. For $\kappa$-phase BEDT-TTF salts the approximation is
that the charge carriers are predominantly localized on the dimers
which are weakly interacting with each other. Although this
implies an insulating system, with the response of delocalized
charge carriers parameterized separately, electron-electron
correlations such as the onsite Coulomb repulsion can be included
explicitly \cite{Rice_PRB1979}.

\subsection{The dimer charge-oscillation model.}
This subsection outlines the derivation of the dimer
charge-oscillation model \cite{Rice_SSC1979,Rice_PRB1979}. Not
only does this aid discussion relating to the validity of applying
this model to $\kappa$-(BEDT-TTF)$_{2}$Cu(SCN)$_{2}$, but it also
lends theoretical weight to the concept of electron coupling to
the antisymmetric combinations of molecular vibrations and not to
the symmetric combinations. This concept is the key to comparing
Raman and infrared data in order to probe the electron-phonon
coupling.

Each dimer unit is modelled by the following Hamiltonian
($\hbar=1$) \cite{Rice_SSC1979,Rice_PRB1979},
\begin{equation}
H = H_{\rm e} + H_{\nu}
+\sum_{\alpha,j}n_{j}g_{\alpha}Q_{\alpha,j}. \label{DCH1}
\end{equation}
The first two terms describe, respectively, the electronic states
and the molecular vibrations in the absence of vibrational
coupling; $j$ labels the molecule and $\alpha$ the normal mode of
vibration. Rice's formulation includes only the totally symmetric
($A_{g}$) molecular modes. However, later theories have extended
the theory to internal modes of other symmetries
\cite{Kozlov_SM1995}.

The electronic term, $H_{\rm e}$, is of the form,
\begin{equation}
H_{\rm e} = \sum_{j=1:2}E_{0}n_{j} + V, \label{DCHe1}
\end{equation}
where $n_{j} = \sum_{\sigma}c^{\dag}_{j,\sigma}c_{j,\sigma}$ is
the occupation number operator for the molecular orbital of energy
$E_{0}$. $c^{\dag}_{j,\sigma}$ and $c_{j,\sigma}$ are the Fermion
creation and annihilation operators for an electron of spin
$\sigma$ \cite{Inkson_1984,Haken_1976}. The exact form of the
potential term, $V$, is arbitrary.

The phonon term, $H_{\nu}$, is of the conventional form
for a
harmonic oscillator of frequency $\omega_{\alpha}$
\cite{Inkson_1984,Rice_PRB1979,Haken_1976},
\begin{equation}
H_{\nu} =
\sum_{j,\alpha}\omega_{\alpha}(b^{\dag}_{j,\alpha}b_{j,\alpha} +
\frac{1}{2}), \label{DCHnu1}
\end{equation}
where $b^{\dag}_{j,\alpha}$ and $b_{j,\alpha}$ are the Boson
creation and annihilation operators. Their sum is the displacement
operator for the $\alpha^{\rm th}$ normal mode and their
difference is its momentum operator \cite{Haken_1976}, {\it i.e.}
\begin{equation}
Q_{j,\alpha} \propto (b^{\dag}_{j,\alpha} + b_{j,\alpha}).
\end{equation}
Because the dimer has been chosen as the base unit for this system
it is appropriate to introduce dimer-mode coordinates
\begin{equation}
s_{\alpha} = \frac{1}{\sqrt{2}}(Q_{1,\alpha} + Q_{2,\alpha})
\end{equation}
and
\begin{equation}
q_{\alpha} = \frac{1}{\sqrt{2}}(Q_{1,\alpha} - Q_{2,\alpha})
\end{equation}
which are symmetric and antisymmetric combinations of the
molecular-mode coordinates respectively.

The third term in the Hamiltonian is a linear perturbation arising
from the electron-molecular vibration coupling. Commutation of the
dimer-mode coordinates with the system Hamiltonian yields two
equations of motion for the $\alpha^{\rm th}$ molecular mode
\cite{Rice_PRB1979,Pines_1966}
\begin{equation}
\langle\ddot{s}_{\alpha}\rangle + \omega^{2}_{\alpha}\langle
s_{\alpha}\rangle = -\sqrt{2}g_{\alpha}\omega_{\alpha}\langle
n_{1}+n_{2}\rangle \label{Sequmotion1}
\end{equation}
and
\begin{equation}
\langle\ddot{q}_{\alpha}\rangle + \omega^{2}_{\alpha}\langle
q_{\alpha}\rangle = -\sqrt{2}g_{\alpha}\omega_{\alpha}\langle
n_{1}-n_{2}\rangle. \label{ASeuqmotion1}
\end{equation}

The right-hand side of the equation of motion for the symmetric
combination of molecular modes is dependent on the total charge
density, $\langle N_{\rm tot}\rangle = \langle n_{1} +
n_{2}\rangle$, which for an isolated dimer is constant, {\it i.e.}
the symmetric combination of modes is uncoupled from the electron
system. Their unperturbed frequencies, $\omega_{\alpha}$, are
therefore obtained experimentally from Raman scattering
experiments \cite{Rice_SSC1979}.

The right-hand side of the equation of motion for the
antisymmetric combination of molecular modes is dependent upon the
electric dipole moment of the dimer, $\langle{\bf p}\rangle =
\frac{1}{2}e{\bf a}\langle n_{1}-n_{2}\rangle$, where $e$ is the
electron charge and ${\bf a}$ is a vector perpendicular to the
interface between the two molecules, whose magnitude is of the
order of the intermolecular separation. Therefore the
antisymmetric combination of modes is directly coupled to the
electron dipole moment and is hence infrared active
\cite{Rice_SSC1979}. For a simple two-state electron Hamiltonian
these modes couple to the charge-transfer excitation between the
ground and excited states, $|1\rangle\rightarrow|2\rangle$.

Because the interaction part of the Hamiltonian is of the form
\begin{equation}
H_{i} = \sum_{k} \rho_{k}\varphi_{k},
\end{equation}
where $\varphi$ is a scalar potential, the expectation value of
the operator $\rho$ which characterizes the response of the system
to the perturbation may be calculated using linear response theory
\cite{Kubo_1956,Pines_1966,Pines_1962}. The ground state
expectation value of the Fourier components of the dipole moment,
$\langle {\bf p}(\omega) \rangle$, and hence the complex
conductivity, $\sigma(\omega)$, of the dimer charge-oscillation
system are given by the linear-response function
\cite{Pines_1966,Rice_SSC1979},
\begin{equation}
\sigma(\omega) = -{\rm i}\omega \frac{e^{2}{\bf
a}^{2}}{2\mathcal{V}} \left(\frac{1}{\chi(\omega)} -
\sum_{\alpha}\frac{g^{2}_{\alpha}\omega_{\alpha}}{\omega^2_{\alpha}-\omega^{2}-{\rm
i}\omega\gamma_{\alpha}}\right)^{-1}. \label{dimerCsigma1}
\end{equation}
Here $\mathcal{V}$ is the molecular volume and $\chi(\omega)$ the
reduced charge transfer polarizability;
\begin{equation}
\chi(\omega) = |\langle2|{\bf p}|1\rangle|^{2} \frac{2\omega_{\rm
CT}}{\omega^2_{\rm CT}-\omega^{2}-{\rm i}\omega\Gamma},
\label{dimerCsigma2}
\end{equation}
where $\langle2|{\bf p}|1\rangle$ is the matrix element for the
charge-transfer transition of frequency $\omega_{\rm CT}$. The
electron and phonon damping factors, $\Gamma$ and
$\gamma_{\alpha}$, are introduced phenomenologically to account
for the observed infrared linewidths. It should be noted that this
perturbation will alter the frequency at which the coupled
molecular vibrations are observed in the infrared spectrum; for
$\omega_{\alpha}~>~\omega_{\rm CT}$ their frequencies will be red
shifted \cite{Rice_PRB1979}.

The crux of this subsection is to address how the dimer
charge-oscillation model may be used to extract the
electron-phonon coupling constants by means of experiment. Several
approaches have been proposed to reduce the possible parameter
space encountered when fitting infrared reflectance data with this
model. It has been noted \cite{Yartsev_IJMPB1998} that the inverse
of the real part of the conductivity, calculated using equation
\ref{dimerCsigma1}, consists of a constant background arising from
the charge-transfer band with peaks occurring at the unperturbed
frequencies of the vibrational modes, $\omega_{\alpha}$. In
principal, if this model accurately describes the data,
Kramers-Kronig analysis could be used to calculate ${\rm
Re}\left(\frac{1}{\sigma(\omega)}\right)$ and thus determine the
bare-mode frequencies. However, because of uncertainties in the
Kramers-Kronig procedure \cite{Klehe_JPCM2000} Raman measurement
of the unperturbed frequencies are more reliable.

Analysis of the poles and zeros of the response function
(Equation~\ref{dimerCsigma1}) leads to the following approximate
relation between the observed frequency of the infrared phonon
bands, $\Omega_{\alpha}$, their unperturbed frequencies,
$\omega_{\alpha}$, the frequency of the coupling charge-transfer
band, $\omega_{\rm CT}$, and the dimensionless electron molecular
vibration coupling constants, $\lambda_{\alpha}$
\cite{Sugano_PRB1989,Bozio_JCP1982};
\begin{equation}
\frac{\omega_{\alpha}^{2}-\Omega_{\alpha}^{2}}{\omega_{\alpha}^{2}}
= \lambda_{\alpha}\frac{\omega_{\rm CT}^{2}}{\omega_{\rm
CT}^{2}-\omega_{\alpha}^{2}}. \label{DCapprox}
\end{equation}
This expression is valid if the bare frequencies of the modes are
well separated from each other or $\lambda_{\alpha}$ sufficiently
small so that each mode may be treated individually. For
$\kappa$-(BEDT-TTF)$_{2}$Cu(SCN)$_{2}$ the two observed modes with
a finite $\lambda_{\alpha}$ (the C=C and C-S modes) are well
separated from each other. Note that the C=C mode overlaps with
two C-H modes as discussed in Section~\ref{modesS}; however the
lack of spectral weight associated with the C-H
modes~\cite{McDonald_SM2001} indicates that their electron-phonon
coupling is negligable in this case. Within the dimer charge
oscillation model $\lambda_{\alpha}$ is related to $g_{\alpha}$
via \cite{Painelli_SSC1984}
\begin{equation}
\lambda_{\alpha} = \frac{2g^{2}_{\alpha}}{\omega_{\alpha}\omega_{\rm
CT}}.
\end{equation}

\subsection{\label{LDC} The electron-phonon coupling constant and its
 pressure derivative derived from the dimer charge-oscillation model.}
Equation~\ref{DCapprox} is used to calculate an electron-phonon
coupling constant, $\lambda_{\alpha}$, for each mode from its
Raman and infrared frequencies, $\omega_{\alpha}$ and
$\Omega_{\alpha}$, given knowledge of the energy, $\omega_{\rm
CT}$, of the coupling charge-transfer band
\cite{Sugano_PRB1989,Bozio_JCP1982} (see Table~\ref{Ltable}). The
dominant source of error in this calculation is due to the width
of the charge-transfer band. There also exists the possibility of
coupling to two different charge-transfer transitions. The C=C
mode, being an antisymmetric combination of A$_{g}$ molecular
modes, couples to the intradimer charge transfer
\cite{Kozlov_SM1995}. It has been calculated, \cite{Kozlov_SM1995}
however, that the antiphase combination of B$_{3g}$ molecular
modes couples to charge transfer perpendicular to the intradimer
direction. This suggests that the C-S mode couples to transitions
between the lower and upper branches of the same band, not
interband electronic transitions. However, it should be noted that
the electron-phonon coupling constants for the C-S mode have been
calculated using the same $\omega_{\rm CT}$ as the C=C mode
because it is impossible to distinguish the contributions to the
infrared spectrum from different transitions.

The pressure derivative of Equation~\ref{DCapprox},
\[
\frac{{\rm d}\ln\lambda_{\alpha}}{{\rm d}P} =
2\frac{\Omega_{\alpha}^{2}}{\omega_{\alpha}^{2}-\Omega_{\alpha}^{2}}\left[\frac{{\rm
d}\ln\omega_{\alpha}}{{\rm d}P}-\frac{{\rm
d}\ln\Omega_{\alpha}}{{\rm d}P}\right]
\]
\begin{equation}
+ 2\frac{\omega_{\alpha}^{2}}{\omega_{\rm
CT}^{2}-\omega_{\alpha}^{2}}\left[\frac{{\rm d}\ln\omega_{\rm
CT}}{{\rm d}P}-\frac{{\rm d}\ln\omega_{\alpha}}{{\rm d}P}\right],
\label{lamdir}
\end{equation}
has a weaker dependence on the value of $\omega_{\rm CT}$. However
the dominant source of error in Equation~\ref{lamdir} arises from
the difficulty in determining an accurate pressure dependence of
the broad charge-transfer band. Fitting the infrared reflectance
data with a simple Drude-Lorentz model~\cite{Klehe_JPCM2000}
provides an estimate for the pressure dependance of $\omega_{\rm
CT}$. Using a value of 4$\pm$4~\%GPa$^{-1}$ for both {\it b}- and
{\it c}-axes \cite{Klehe_JPCM2000}, results in an error in the
pressure derivative of the coupling constant for the C=C mode of
$\pm$3.5~\%GPa$^{-1}$ for the {\it b}-axis and $\pm$5~\%GPa$^{-1}$
for the {\it c}-axis. The error in the pressure derivative of the
coupling constant for the C-S mode is larger (8~\%GPa$^{-1}$ for
the {\it b}-axis and 28~\%GPa$^{-1}$ for the {\it c}-axis) because
$\lambda_{\rm CS}$ is so small.
\begin{table}[ht]
\begin{center}
\begin{tabular}{ccccc} \hline \hline
Mode & $\lambda_{b}$ & $\frac{{\rm d}\ln\lambda_{b}}{{\rm d}P}$ &
$\lambda_{c}$ & $\frac{{\rm d}\ln\lambda_{c}}{{\rm d}P}$\\
 & & \%GPa$^{-1}$ & & \%GPa$^{-1}$\\
  \hline\hline
   $\omega_{\rm CS}$ & 0.01(1) & 47.9 & 0.02(1) & -11.1\\
    $\omega_{\rm CC}$ & 0.17(1) & -11.9 &  0.20(1) & -17.3\\
     $\omega_{\rm anion1}$ & 0 & 0 & 0 & 0\\
     $\omega_{\rm anion2}$ & $\cdots\cdots$ & $\cdots\cdots$ & 0 & 0\\
      \hline
\end{tabular}
\caption{dimensionless electron-phonon coupling constants and
their pressure derivatives calculated using the dimer
charge-oscillation model. Note in this table the $\omega$
subscript referes to the mode and the $\lambda$ subscript to the
polarization direction.} \label{Ltable}
\end{center}
\end{table}

Because the dimer charge-oscillation model is formulated for a
localized system, $\lambda_{\alpha}$ calculated in this manner
parameterizes the strength of interaction between the phonons and
the intradimer charge transfer. On the other hand the
electron-phonon coupling constant associated with phonon-mediated
superconductivity parameterizes the strength of interaction
between the phonons and the delocalized charge carriers. Therefore
to use this model to draw any conclusions regarding the
superconducting mechanism requires the assumption that the
electron-phonon coupling for the intradimer charge transfer will,
to first order, have the same pressure dependence as the
electron-phonon coupling for the interdimer charge transfer. This
assumption is reasonable because it is the same molecular orbitals
that are responsible for intra- and interdimer wave function
overlap. At the $\Gamma$-point (the long wavelength limit)
neighboring dimers vibrate in phase~\cite{Dove_1993} and as a
result, an antiphase combination of molecular vibrations that
modulates the intradimer wavefunction overlap will also modulate
the interdimer wavefunction overlap.

\subsection{\label{PPmodel}The phase-phonon model.}
Phase-phonon theory essentially describes the same phenomenon as
the dimer charge-oscillation theory (see equation \ref{DCH1}).
However, because it is formulated in reciprocal space the model is
easily extended to metallic systems. This model is not just
limited to the dimer unit, but attributes the infrared activity of
the coupled phonon modes to phase oscillations of the spatial
charge density, induced by the electron-phonon interaction
\cite{Rice_PRL1976}.

The system Hamiltonian is similar to that used in the dimer
charge-oscillation theory, and includes an electronic term,
$H_{\rm e}$, a phonon term, $H_{\nu}$, and a linear coupling term,
$H_{\rm i}$.
\begin{equation}
H = H_{\rm e} + H_{\nu}
+\frac{1}{\sqrt{\mathtt{N}}}\sum_{\alpha,q}g_{\alpha}Q_{\alpha}(q)\rho_{-q}.
\label{DPP1}
\end{equation}
The electronic terms has essentially the same form as
(\ref{DCHe1}),
\begin{equation}
H_{\rm e} = \sum_{k,\sigma}E_{k}c^{\dag}_{k,\sigma}c_{k,\sigma} +
V(\rho_{q_{0}}+\rho_{-q_{0}}), \label{DPPe2}
\end{equation}
except that the operators create or annihilate a particle in a
$k$-state, not at a real space location, {\it i.e.} it describes a
system of conduction electrons moving in a weak periodic
potential, $V$, of wavevector $q_{0}$. The operator $\rho_{q} =
\sum_{k}c^{\dag}_{k+q}c_{k}$ creates an electronic-density
fluctuation of wavevector $q$.

The phonon term is identical to that used in the dimer-charge
model except the phonon operators also act on $k$-states. The
electron-molecular vibration coupling is included as a linear
perturbation, in this case linking the Fourier transform of the
mode displacement vector, $Q_{\alpha}(q) \propto
(b_{\alpha}(q)+b^{\dag}_{\alpha}(-q))$, to an electronic density
fluctuation of the same wavevector, $q$.
\begin{equation}
H_{i} = \frac{1}{\sqrt{\mathtt{N}}}\sum_{\alpha,q,k}g_{\alpha}
c^{\dag}_{k+q}c_{k} (b_{\alpha}(q)+b^{\dag}_{\alpha}(-q)).
\end{equation}
$\mathtt{N}$ is the density of conduction electrons.

For the case initially considered by Rice, \cite{Rice_PRL1976}
$q_{0} = 2k_{F}$ (twice the Fermi wavevector) such that the
periodic potential induces the conduction electrons to condense
into a charge-density-wave state. However, the derivation that
follows is equally applicable to a metallic system. The
frequency-dependent conductivity for an insulating system is due
simply to interband electronic transitions \cite{Lee_SSC1974},
\begin{equation}
\sigma(\omega) = \frac{\omega_{\rm p}^{2}}{{\rm i}\omega}\left[f(x)-f(0)\right], \label{sigPP1}
\end{equation}
where the plasma frequency, $\omega_{\rm p}$, parameterizes the
band curvatures. $f(0) = 1$ and
\begin{equation}
f(x) = \frac{\left[{\rm i}\pi +
\ln\left(\frac{1-S}{1+S}\right)\right]}{2Sx^{2}}
\end{equation}
where $S = \sqrt{1-\frac{1}{x^{2}}}$ and $x =
\frac{\omega}{2\Delta_{0}}$ with $\Delta_{0}$ equal to the band
gap. Including the electron-phonon coupling modulates the onsite
electron energy and hence the band gap, so that $\Delta =
\Delta_{0} + \sum_{\alpha}\Delta_{\alpha} e^{{\rm
i}\phi_{\alpha}}$, where $\Delta_{\alpha}$ and $\phi_{\alpha}$ are
the amplitude and phase respectively of the distortion potentials
which are proportional to $g_{\alpha}$. The single electron
contributions to $\sigma(\omega)$ are of the same form as the
uncoupled case but with $x = \frac{\omega}{2\Delta}$. In addition
to the single-electron contributions there are collective
contributions associated with oscillations in the phases,
$\phi_{\alpha}$, of the combined lattice-charge distortions.
Collective modes associated with oscillations in amplitude,
$\Delta_{\alpha}$, also occur: however, they preserve dipole
moment and hence do not contribute to the dielectric response
\cite{Rice_PRL1976,Rice_SSC1977}.

Including the collective mode contribution, the
frequency-dependent conductivity is given by~\cite{Rice_PRL1976}
\begin{equation}
\label{sigPP}
\sigma(\omega) = - \frac{\omega_{\rm p}^{2}}{{\rm i}\omega}
\left[f(x)-f(0)-x^{2}(f(x))^{2}\lambda D_{\phi}(x)\right].
\end{equation}
$D_{\phi}$ is a phonon-like propagator for the phase oscillations given by
\begin{equation}
\frac{1}{D_{\phi}} = \frac{1}{D_{0}} + 1 - \frac{\Delta_{0}}{\Delta}
+ \lambda x^{2} f(x).
\end{equation}
$D_{0}$ is the same single-phonon propagator used in the dimer charge-oscillation model,
\begin{equation}
D_{0} = -\sum_{\alpha}\frac{\lambda_{\alpha}}{\lambda}
\frac{\omega_{\alpha}^{2}}{\omega^2_{\alpha}-\omega^{2}-{\rm
i}\omega\gamma_{\alpha}}.
\end{equation}
In the phase-phonon model the dimensionless coupling constant,
$\lambda$ is related to $g_{\alpha}$ via
\begin{equation}
\lambda_{\alpha} =
\frac{\mathtt{N}(k_{F})g_{\alpha}^{2}}{\omega_{\alpha}},
\end{equation}
where $\mathtt{N}(k_{F})$ is the density of electron states at the
Fermi-surface and the total electron-phonon coupling constant,
$\lambda = \sum_{\alpha}\lambda_{\alpha}$
\cite{Yartsev_IJMPB1998}.

It is necessary to include an electronic damping term, $\Gamma$,
to account for the finite electron lifetime if the model is to be
applied to a metallic system, {\it i.e.} when a large plasma
frequency is required to account for the intraband processes. In
this case $f(x)$ is replaced by $f(x+{\rm i}\Gamma)$
\cite{Fenton_SSC1983}.

For $\omega_{\alpha} < 2\Delta$ decay of the collective mode via excitation of a real electron-hole pair is energetically impossible. As a result each mode contributes a sharp absorption band to the infrared spectrum whose width is determined solely by the lineshape of the uncoupled phonon mode. For $\omega_{\alpha} > 2\Delta$ the collective modes become damped via electron-hole excitation, appearing as indentations in the electronic background \cite{Rice_SSC1977}.

\subsection{The electron-phonon coupling constant and its pressure derivative derived from the phase-phonon model.}
The electron-phonon coupling constant that occurs in the
phase-phonon model parameterizes the degree to which each
molecular vibration modulates the energy of the electronic band
structure, {\it i.e.} it parameterizes both the inter- and
intraband electron-phonon coupling.
\begin{figure*}[htb]
\includegraphics[width=12cm]{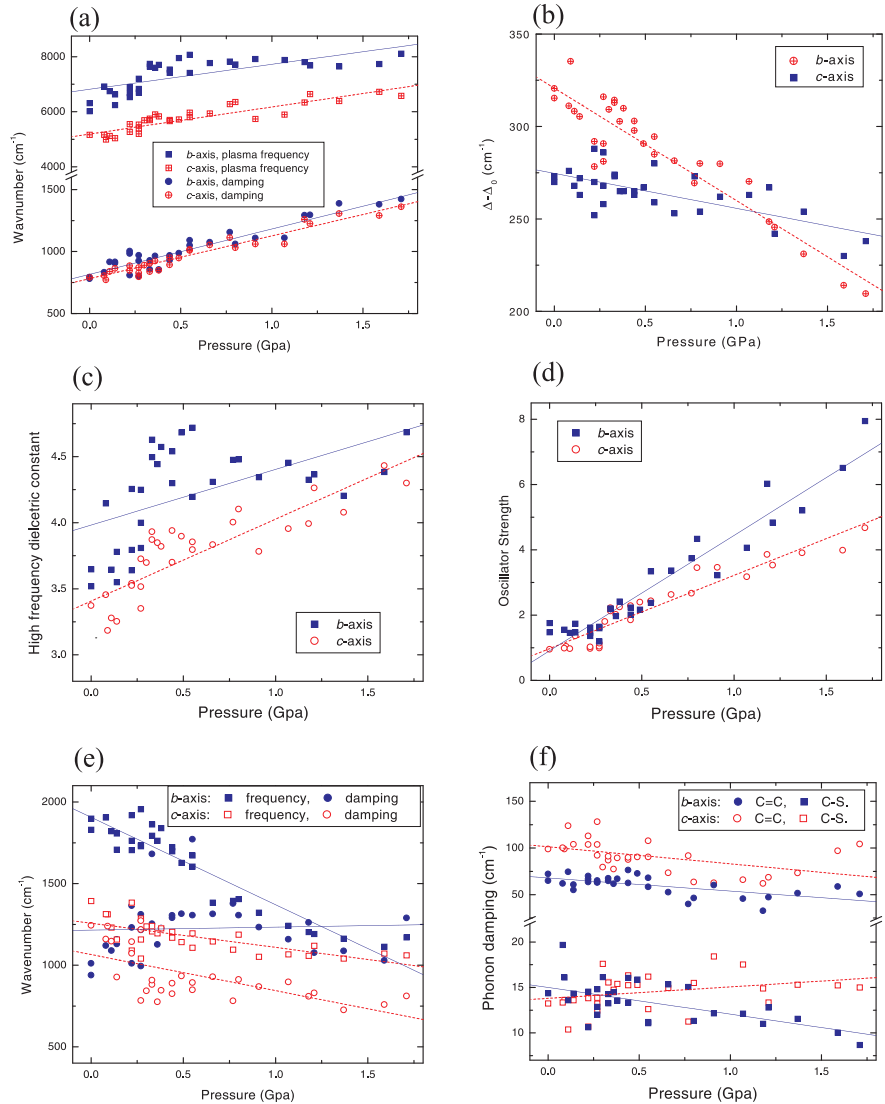}
\centering \caption{The pressure dependence of the free parameters of the phase-phonon model. Note that the linear fits indicate the first order pressure dependence used for the batch fitting procedure. Solid lines and symbols are used for polarization parallel to the {\it b}-axis and dashed lines and hollow symbols for polarization parallel to the {\it c}-axis. (a) the plasma frequency and electronic damping. (b) the difference between the perturbed and unperturbed interband transition frequency. (c) the high frequency dielectric constant. (d) the strength of the Lorentzian oscillator. (e) the frequency and damping of the Lorentzian oscillator. (f) the damping of the C-S and C=C modes.} \label{PPpargraph1}
\end{figure*}
\begin{figure*}[htb]
\includegraphics[width=14cm]{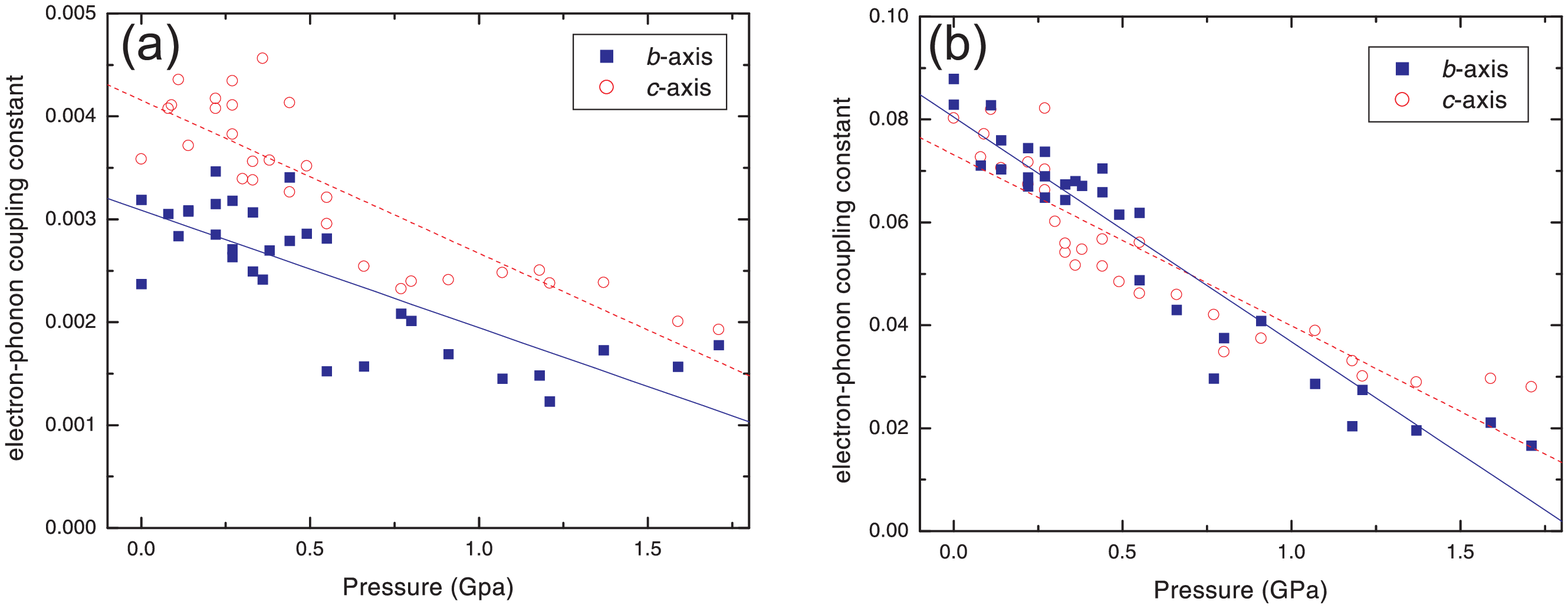}
\centering \caption{The pressure dependence of the electron-phonon coupling constant. Solid squares and lines for polarization parallel to the {\it b}-axis and hollow circles and dashed lines for polarization parallel to the {\it c}-axis. (a) for the C-S mode and (b) for the C=C mode.} \label{PPlambdagraph1}
\end{figure*}
\begin{table*}[htb]
\begin{center}
\begin{tabular}{ccccc} \hline \hline
Parameter & b-axis & value & c-axis & value\\
 & cm$^{-1}$ & +$\%$GPa$^{-1}$ & cm$^{-1}$ & +$\%$GPa$^{-1}$\\
  \hline\hline
$\omega_{p}$ & (6816$\pm$120) & +(13$\pm$2) & (5190$\pm$66) & +(19$\pm$2)\\ $\Gamma$ & (815$\pm$16) & +(45$\pm$3) & (781$\pm$12) & +(45$\pm$3)\\ $\Delta_{0}$ & (1132$\pm$13) & +(10$\pm$2) & (743$\pm$7) & +(17$\pm$2)\\ $\Delta$ & (1407$\pm$14) & +(6$\pm$2) & (1064$\pm$8) & +(6$\pm$1)\\ $\Delta-\Delta_{0}$ & (275$\pm$3) & -(7$\pm$2) & (320$\pm$3) & -(19$\pm$2)\\ $\gamma_{\rm CS}$ & (15$\pm$0.5) & -(26$\pm$7) & (13$\pm$0.5) & +(8$\pm$8)\\ $\gamma_{\rm CC}$ & (68$\pm$2) & -(21$\pm$5) & (101$\pm$5) & -(18$\pm$6)\\ $\omega_{e}$ & (1905$\pm$25) & -(28$\pm$2) & (1257$\pm$21) & -(12$\pm$2)\\ $\gamma_{e}$ & (1214$\pm$51) & +(2$\pm$6) & (1065$\pm$38) & -(21$\pm$2)\\
\hline
&  & +$\%$GPa$^{-1}$ &  & +$\%$GPa$^{-1}$\\
\hline $\Delta\varepsilon_{e}$ & (0.90$\pm$0.15) & +(388$\pm$22) & (1.00$\pm$0.10) & +(272$\pm$15)\\
$\varepsilon_{\infty}$ & (3.9$\pm$0.1) & +(11$\pm$3) & (3.4$\pm$0.1) & +(18$\pm$3)\\
$\lambda_{\rm CS}$ & (3$\times10^{-3}\pm$1$\times10^{-4}$) & -(32$\pm$5) & (3$\times10^{-3}\pm$1$\times10^{-4}$) & -(25$\pm$4)\\
$\lambda_{\rm CC}$ & (0.080$\pm$0.002) & -(55$\pm$4) & (0.073$\pm$0.002) & -(44$\pm$3)\\
\hline
\end{tabular}
\caption{Phase phonon model parameters and their first order pressure
shifts.} \label{PPtable1}
\end{center}
\end{table*}

To calculate $\lambda_{\alpha}$ for each mode the full
diamond/sample IR reflectance spectrum~\cite{Klehe_JPCM2000} is
fitted at each pressure-point. The model function for this fit
consists of the phase-phonon model plus a high frequency
contribution to the dielectric constant, $\varepsilon_{\infty}$,
and a highly damped Lorentzian oscillator to account for the
anomalous electronic damping at room temperature~\cite{endnote2}.
The plasma frequency, electronic damping and perturbed and
unperturbed interband transition frequencies are treated as free
parameters. As with a simple Drude response, the plasma frequency
parameterizes the density of states and band curvature, but in
this case the electronic damping controls the relative spectral
weights of the inter- and intraband processes.

Four phase-phonon collective modes ($\alpha = 1\rightarrow4$) are
used to model the spectrum, one for the C-S mode and three for the
C=C mode and its Fermi resonance with the C-H modes. It was found
that the line shape of the Fermi resonance could be reproduced in
an analogous manner to the Green's function model used in
\cite{McDonald_SM2001}, {\it i.e.} with one strongly coupled C=C
mode and two C-H modes with negligible electron-phonon coupling
(see the inset of Figures~\ref{PPfitb} and \ref{PPfitc}). The
pressure dependence of the modes' Raman frequencies and the
pressure dependence of the dips from the Green's function model
were used to fix the frequency of the collective modes. Their
damping and coupling strengths were left as free parameters,
however.

In this model the degree of mode softening associated with the
electron-phonon coupling is predominantly parameterized by the
perturbation to the interband transition, {\it i.e.} the
difference between the perturbed and unperturbed transition
frequencies. The coupling strength, $\lambda_{\alpha}$, controls
the mode softening to a lesser extent, mainly parameterizing the
spectral weight associated with each mode. Because of this,
correctly determining $\lambda$ relies on accurately modelling the
background reflectance in the spectral region containing the
modes. To this end, it was found that including the heavily damped
Lorentzian to help model this material's anomalous Drude response
drastically reduced the scatter in the $\lambda$ data. The three
parameters describing the Lorentzian, $\omega_{e}$,$\gamma_{e}$
and $\Delta\varepsilon_{e}$, were also free.

The lineshape of the anion modes is most accurately reproduced
using uncoupled Lorentzian oscillators. Owing to their lack of
electron coupling and their limited contribution to the spectrum,
parameters describing them were not included in this model.
Figures~\ref{PPfitb} and \ref{PPfitc} indicate the quality of the
fits for a representative selection of pressures.

Manually fitting 31 spectra, each consisting of 3300 data points,
with an expression containing 12 free parameters, for each
polarization, is both immensely time consuming and subject to
systematic human error. To avoid this a batch fitting procedure
was employed. It consists of accurately fitting a limited number
of data sets to determine approximate trends for the free
parameters. These are then used as the the starting conditions for
an automatic least-squares fitting procedure of all data sets.
Each fit is then checked visually for acceptance of its
parameters.

The value of the free parameters and their first order pressure
shifts extracted from Figures~\ref{PPpargraph1} and
\ref{PPlambdagraph1} are contained in Table~\ref{PPtable1}. The
difference in the values of the coupling constants, $\lambda$,
obtained from the phase-phonon model (TABLE \ref{PPtable1}) and
the dimer charge-oscillation model(TABLE \ref{Ltable}) are
predominantly due to the different values for the interband
transition used in each calculation.

\section{Discussion.}
As well as vibrational features the room temperature IR spectrum
of $\kappa$-(BEDT-TTF)$_{2}$Cu(SCN)$_{2}$ contains broad
electronic features; a heavly damped Drude-like response in the
far-infrared and a mid-infrared (MIR) `hump'. Regarding the MIR
`hump' the literature only agrees to the extent that it originates
from excitation across a gap in the electronic
spectrum~\cite{Kornelsen_PRB1991,Merino_CCMP1998,Wang_PC2000}. The
simplest picture is to assign it to a gap in the single electron
band structure \cite{Kornelsen_PRB1991}. This model is improved by
including electron-electron interaction effects
\cite{Merino_CCMP1998}. In this case the `hump' originates from a
combination of the single electron band gap and the onsite
correlation energy. Which is considered dominant, and which the
perturbation, depends upon how correlated the system is believed
to be. Analogously, inclusion of the electron-phonon interaction
modifies the electronic spectrum. The phase-phonon model includes
this explicitly as a relatively small perturbation to the band
gap. In the limit of large electron-phonon coupling the
strain-field is considered to localize the charge carrier
\cite{Wang_PC2000}, in which case the small polaron binding energy
will be the band gap dressed by the electron-phonon interaction.
All these theories share the common assumption that IR active
interband electronic processes are subject to remormalization by
many-body effects. In the following subsection we do not attempt
to identify the origin of the renormalizations, but present
evidence of their influence on the pressure dependence of the
electronic contributions to the IR spectrum.

\subsection{\label{M*}The pressure dependence of the carrier effective mass.}
This subsection focuses on the pressure dependence of the carrier
effective mass extracted from modeling with the phase-phonon
theory.  The plasma frequency, $\omega_{\rm p}$, parameterizes the
charge carrier density, $\mathtt{N}$, and effective mass, here
denoted $m^{*}_{\rm pp}$,
\begin{equation}
\omega_{\rm p}^{2} = \frac{\mathtt{N}e^{2}}{\epsilon_{0}m^{*}_{\rm
pp}}.
\end{equation}
For our actual pressure range the band filling is independent of
pressure and the carrier density only has a small pressure
dependence arising from the reduction in unit cell parameters
\cite{Klehe_JPCM2000,Caulfield_JPCM1994}.

For a plasma frequency dominated by intraband processes the
pressure dependence of $m^{*}_{\rm pp}$ arises solely from the
pressure dependence of the band parameters
\cite{Legget_AoP1968,Klehe_JPCM2000}, {\it i.e.} the increase in
band width and curvature as the wave function overlap increases.
We argue that if the plasma frequency also parameterizes the
interband processes it will contain contributions from
quasiparticle mass renormalization. The first point to note (see
Section~\ref{PPmodel}) is that the plasma frequency in the phase
phonon model not only describes screening ability of the free
carriers (the intraband response) but also the density of states
and band curvatures associated with intraband absorption. Thus
following the preceeding logic the mass $m^{*}_{\rm pp}$ will to
some extent be renormalized by many-body
effects~\cite{Legget_AoP1968}.

The second point to note is that using the pressure dependence of
the Raman modes to constrain the mode frequencies for the
phase-phonon model significantly  reduces the parameter space
encountered when modeling the IR response of this system. This is
not only because it eliminates the mode frequencies as free
parameters, but because the softening of the modes is dependent
upon the perturbation to the band gap, $\Delta - \Delta_{0}$. In
this manner the mode frequencies provide further constraint to the
parameters describing the electronic response. The parameters
describing the electronic response are thus believed to have
greater validity than those obtained from a simple Drude-Lorentz
fit~\cite{Klehe_JPCM2000}.

\begin{figure}[ht]
\includegraphics[width=9cm]{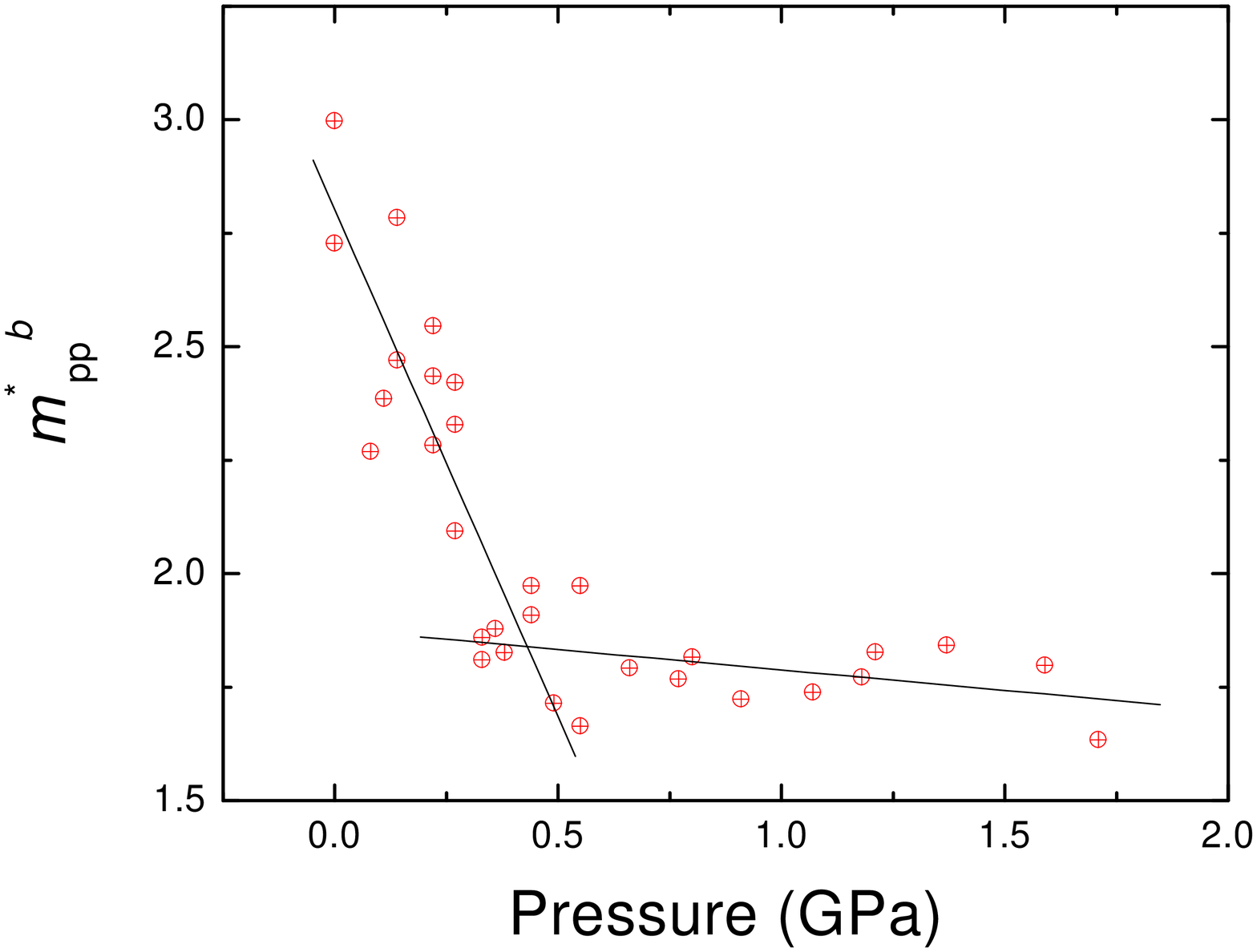}
\centering \caption{The mass $m^{*~b}_{\rm pp}$ for light
polarized parallel to the {\it b}-axis, deduced from the
phase-phonon fit to the optical data. For $P<0.5 {\rm PGa}$, ${\rm
d}m^{*~b}_{\rm pp}/dP\sim-2.2{\rm m}_{\rm e}$ and d$\ln{
m^{*~b}_{\rm pp}}/dP\sim-80\%/{\rm GPa}$. For $P>0.5 {\rm PGa}$,
${\rm d}m^{*~b}_{\rm pp}/dP\sim-0.9{\rm m}_{\rm e}$ and d$\ln{
m^{*~b}_{\rm pp}}/dP\sim-5\%/{\rm GPa}$. } \label{mbstar1}
\end{figure}

\begin{figure}[ht]
\includegraphics[width=9cm]{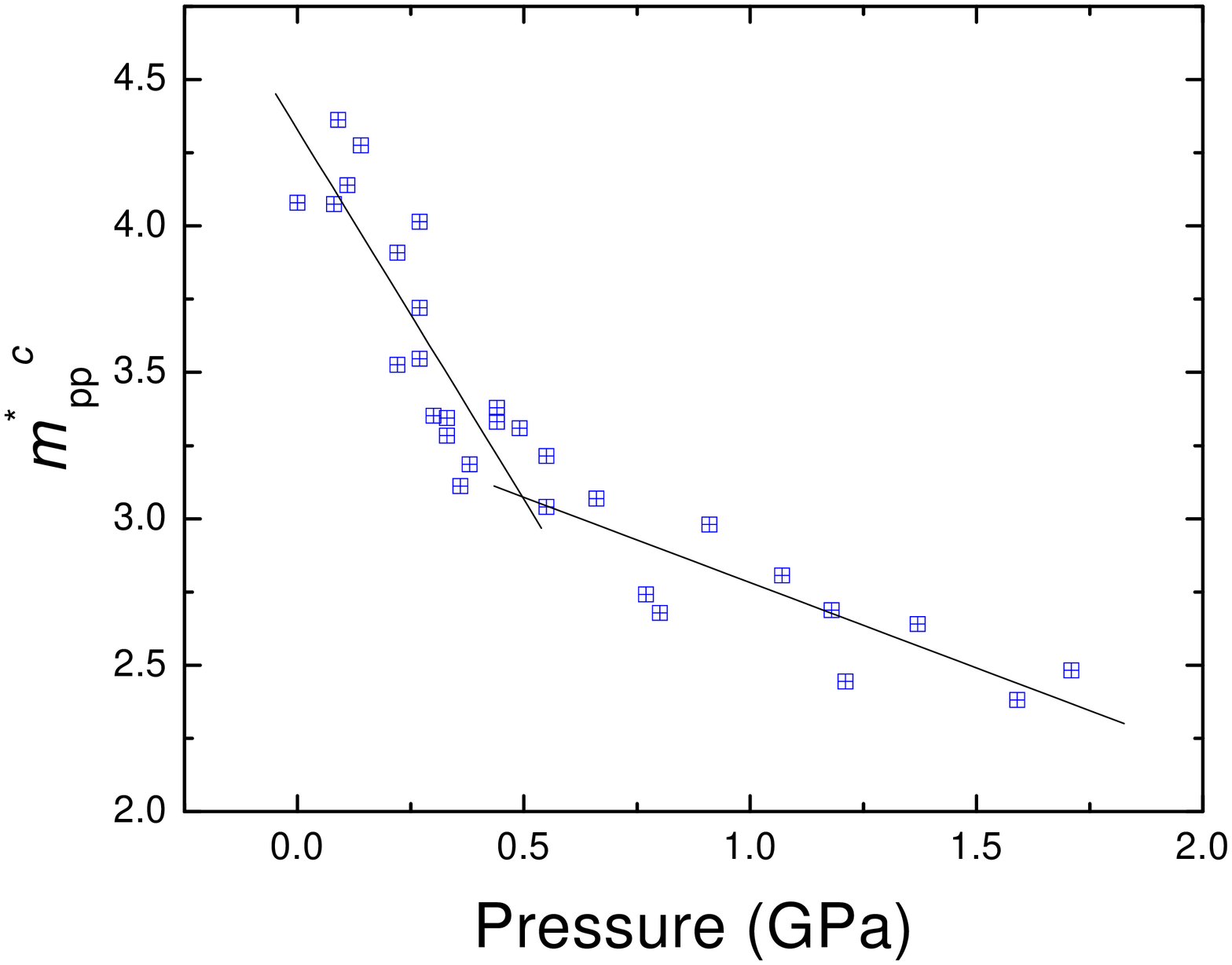}
\centering \caption{The mass $m^{*~c}_{\rm pp}$ for light
polarized parallel to the {\it c}-axis, deduced from the
phase-phonon fit to the optical data. For $P<0.5 {\rm PGa}$, ${\rm
d}m^{*~c}_{\rm pp}/dP\sim-2.5{\rm m}_{\rm e}$ and d$\ln{
m^{*~c}_{\rm pp}}/dP\sim-58\%/{\rm GPa}$. For $P>0.5 {\rm PGa}$,
${\rm d}m^{*~c}_{\rm pp}/dP\sim-0.6{\rm m}_{\rm e}$ and d$\ln{
m^{*~c}_{\rm pp}}/dP\sim-17\%/{\rm GPa}$.} \label{mcstar1}
\end{figure}

The pressure dependence of the effective masses $m^{*}_{\rm pp}$
obtained using the phase-phonon theory is shown in
Figures~\ref{mbstar1}~and~\ref{mcstar1}.
For both polarization directions  $m^{*}_{\rm pp}$  decreases with
the application of pressure. For both polarizations this decrease
is clearly nonlinear and this is highlighted by linear fits to the
low ($<0.5$~GPa) and high pressure regions ($>0.5$~GPa)(see
Fig.~\ref{mbstar1}~and~\ref{mcstar1}). These pressure ranges were
chosen because the {\it b}-axis $m^{*~b}_{\rm pp}$ clearly
exhibits a change of slope at $0.5$~GPa. The trend in $m^{*~c}_{\rm
pp}$ for polarization parallel to the {\it c}-axis is far less
clear-cut. Band structure calculations giving the pressure
dependence of the bare band mass~\cite{McDonald_?2002} indicate a
sublinear behaviour for its pressure dependence. However this is a
much smaller correction to linear behavior than observed here and
fails to reproduce the change in slope at $0.5$~GPa. The
coincidence of `kinks' in the pressure dependence of $m^{*}_{\rm
pp}$ obtained here and $m^{*}$ from magnetic quantum oscillation
data~\cite{Caulfield_JPCM1994} is thus an indication that the
phase-phonon fitting procedure is sensitive to many-body effects.
It is these renormalizations which are known to strongly influence
superconductivity in this material~\cite{Klehe_JPCM2000,
Caulfield_JPCM1994}.

\subsection{\label{TcSS}Implications for the superconducting pairing mechanism.}
It has been shown previously \cite{Yamaji_SSC1998} that the usual
electron-acoustic phonon interaction mechanism is unable to
account for the magnitude of the electron-phonon coupling constant
or the large pressure dependence of $T_{\rm c}$ in the BEDT-TTF
superconductors. A further refinement \cite{Yamaji_SSC1998}
includes the attractive interaction mediated by the A$_{g}$
molecular modes, with the total electron-phonon coupling constant,
$\lambda_{\rm TOT}$, given by a Yamaji sum over the individual
A$_{g}$ molecular modes. The energy scale for the interaction is
set by the Debye frequency, $\Theta$
\cite{Yamaji_SSC1998,Faulhaber_SM1993}. Caulfield {\it et al.}
\cite{Caulfield_JPCM1994} have previously inferred
$\Theta~\approx$~40cm$^{-1}$ by fitting the effective mass
dependence of $T_{\rm c}$ with a linearised Eliashberg equation
using an Einstein density of phonon states, $\delta(\Theta)$. High
and low temperature specific heat measurements
\cite{Andraka_PRB1998,Fortune_SM1999} yield values of $\Theta$
ranging from 38cm$^{-1}$ to 140cm$^{-1}$. Calculations of
$\lambda_{\rm TOT}$ \cite{Yamaji_SSC1998} give values ranging from
0.25-0.45 \cite{Drozdova_SM1994,Hill_SM1993,Shumway_PRB1996}.

The weak-coupling BCS formula \cite{Bardeen_PR1957} gives a
satisfactory functional description of the ambient pressure
$T_{\rm c}$ \cite{Sugano_PRB1989} and describes the effective mass
dependence of the superconducting transition temperature
\cite{Caulfield_JPCM1994}.However, the fact that the weak-coupling
BCS expression describes $T_{\rm c}$ versus $m^*$ well
\cite{Caulfield_JPCM1994} should NOT be taken to imply that
  $\kappa$-(BEDT-TTF)$_{2}$Cu(SCN)$_{2}$ is a weak-coupling BCS superconductor.
  The formula is used here as a convenient parameterization which is known to
  describe earlier data well \cite{Caulfield_JPCM1994,Caulfield_SM1995}.

The pressure derivative of the weak-coupling BCS formula provides
a convenient parametrization of $\frac{{\rm d}\ln~T_{\rm c}}{{\rm
d}P}$ in terms of $\frac{{\rm d}\lambda}{{\rm d}P}$,
\begin{equation}
\frac{{\rm d}\ln~T_{\rm c}}{{\rm d}P} = \frac{{\rm d}\ln\Theta}{{\rm
d}P} + \frac{1}{(\lambda-\mu^{*})^{2}}\left[\frac{{\rm
d}\lambda}{{\rm d}P}-\frac{{\rm d}\mu^{*}}{{\rm d}P}\right].
\label{pressTc}
\end{equation}
With
\begin{equation}
\frac{1}{(\lambda-\mu^{*})^{2}} = \left[\ln\left(\frac{T_{\rm
c}}{1.13\Theta}\right)\right]^{2}, \label{pressTc2}
\end{equation}
$\Theta~\approx$~90$\pm$50~cm$^{-1}$
$[$\onlinecite{Caulfield_JPCM1994,Andraka_PRB1998,Fortune_SM1999}$]$
and using the average pressure-induced stiffening of the Raman
active lattice modes of $\approx$~+13~\%GPa$^{-1}$
$[$\onlinecite{McDonald_JPCM2001}$]$ for $\frac{{\rm
d}\ln\Theta}{{\rm d}P}$, the only unknown is the pressure
dependence of the Coulomb pseudopotential, $\frac{{\rm
d}\mu^{*}}{{\rm d}P}$. There is sufficient uncertainty in the
calculations of $\mu^{*}$ that an accurate estimate for its
pressure dependence cannot be made \cite{Lee_SSC1998}. However,
$\mu^{*}$ is known to scale with the density of states at the
Fermi energy~\cite{Lee_SSC1998}, which will decrease with the
application of pressure~\cite{Caulfield_JPCM1994}. Thus if
$\mu^{*}$ is positive~\cite{Shumway_PRB1996} it will decrease with
pressure and omitting $\frac{{\rm d}\mu^{*}}{{\rm d}P}$ from
Equation~\ref{pressTc} will result in an overestimate for the rate
of  suppression of $T_{\rm c}$ with pressure. A negative value of
$\mu^{*}$ indicates that direct (non phonon-mediated)
electron-electron interactions are involved in the
pairing~\cite{Caulfield_JPCM1994,Lee_SSC1998} which are not
measured by these experiments. Consequently $\frac{{\rm
d}\mu^{*}}{{\rm d}P}$ is ignored in Equation~\ref{pressTc}.

Estimates for the pressure dependence of the electron-phonon
coupling constant, based on the vibrations sampled and the two
methods of calculation are used to calculate the right-hand side
of Equation~\ref{pressTc}.  It should be emphasized that the two
models we are employing to estimate the pressure dependence of the
electron-phonon coupling constant lie at opposite limits of our
sample's behaviour, the dimer charge-oscillation model being the
limit of a localized electronic system and the phase-phonon model
the limit of an itinerant system. A conclusion common to both
methods of analysis should thus be a robust one.

\subsubsection{$T_{\rm c}(P)$ from the dimer charge-oscillation model.}
Employing the dimer charge model, $\lambda_{\rm TOT}$ is assumed
to have a pressure dependence similar to that of the strongly
coupled C=C mode, {\it i.e.} of the order of -17~$\%$GPa$^{-1}$,
with an upper limit of -20$\%$GPa$^{-1}$ used in this calculation.
This gives $\frac{{\rm d}\ln~T_{\rm c}}{{\rm
d}P}~\approx$~-40$\pm32~\%$GPa$^{-1}$, which is far from the
experimentally observed value of $\frac{{\rm d}\ln~T_{\rm c}}{{\rm
d}P}~\approx$~-200~\%GPa$^{-1}$
$[$\onlinecite{Caulfield_JPCM1994,Murata_SM1989}$]$.

As can be seen from (\ref{pressTc2}), $\Theta$ logarithmically
scales with the dependence of the pressure derivative of $T_{\rm
c}$ on the pressure derivative of the coupling constant. Thus to
obtain the observed rapid fall of $T_{\rm c}$ with pressure from
only the decrease in the electron-phonon coupling constant
requires $\Theta$ to be of the order of the C=C mode frequency,
$\approx$~1500~cm$^{-1}$. Such a value is inconsistent with the
10~K superconducting temperature and the unconventional isotope
shift observed upon carbon substitution \cite{Schlueter_PC2001}.
Thus, the electron-phonon coupling as extracted from the
dimer-charge-oscillation model cannot be the relevant parameter
for superconductivity in our organic superconductor.

\subsubsection{$T_{\rm c}(P)$ from the phase-phonon model.}
The pressure dependence of the electron-phonon coupling constant
derived from the phase-phonon model provides far less clear-cut
results. In this case $\lambda_{\rm TOT}$ is assumed to have a
pressure dependence of -40$\pm$20~$\%$GPa$^{-1}$, a value
consistent with the phase-phonon calculation of $\lambda$ for all
observed modes. This gives $\frac{{\rm d}\ln~T_{\rm c}}{{\rm
d}P}~\approx$~-90~$\%$GPa$^{-1}$ with the asymmetric error of +150
and -77~$\%$GPa$^{-1}$. The upper bound of this estimate is close
to the experimentally observed value of $\frac{{\rm d}\ln~T_{\rm
c}}{{\rm d}P}~\approx$~-200~\%GPa$^{-1}$
$[$\onlinecite{Caulfield_JPCM1994,Murata_SM1989}$]$. Thus, the
upper estimate of all parameters would be required to explain the
pressure dependence of $T_{\rm c}(P)$ as arising solely due to a
reduction in the electron-phonon coupling constant.

\subsubsection{Comparison.}
Again, it should be stressed that the dimer charge-oscillation
model represents the limit of localized electronic behaviour and
the phase-phonon model represents the limit of itinerant
electronic behaviour. The true properties of
$\kappa$-(BEDT-TTF)$_{2}$Cu(SCN)$_{2}$ under these experimental
conditions are believed to lie somewhere between these extremes.
Both models have similar conclusions , {\it i.e.} $\frac{{\rm
d}\ln~T_{\rm c}}{{\rm d}P}$ cannot be reproduced within weak
coupling BCS theory. The fact that the similar conclusions are
drawn when employing either model suggests that they are robust.

Use of the dimer charge-oscillation model casts considerable doubt
on whether this material is a simple BCS superconductor because
the characteristic energy of the pairing interaction would have to
be of the order of the highest frequency molecular modes, a value
inconsistent with the 10~K superconducting transition temperature.
However this conclusion must be treated with caution due to the
numerous assumptions involved in applying this model of a
localized system to the coupling between the molecular vibrations
and delocalized conduction electrons. Doubt on whether this
material is a simple BCS superconductor is also cast by the
application of the phase-phonon model, because the upper estimate
of all parameters is required to explain the experimentally
observed suppression of $T_{\rm c}$ with pressure.

This indicates that electron-electron interaction may be playing a
significant role in this material's superconducting mechanism. It
is worth noting that there is mounting evidence in support of this
conclusion. Both experimental \cite{Arai_PRB2001,Izawa_PRL2002}
and theoretical \cite{Kuroki_Xarc2001}, studies predict that the
superconducting order parameter in this material has an `exotic'
d-wave symmetry, a property predicted for a superconducting state
whose pairing is mediated by spin fluctuations
\cite{Kuroki_Xarc2001,Annett_CP1995}, {\it i.e.} direct
electron-electron interaction~\cite{Singleton_RPP2000}.

\section{\label{conc}Conclusions.}
Comparison of high-pressure Raman scattering and infrared
reflectivity data  provides an alternative method for probing the
quasiparticle contributions to effective mass enhancement known to
be intimately connected to superconductivity in this
material~\cite{Caulfield_JPCM1994}. It also enables the pressure
dependence of the electron-phonon coupling strength to be
evaluated for modes observed in both spectra. Using the weak
coupling limit of BCS theory the pressure dependence of the
electron-phonon coupling constant is compared to the pressure
dependence of the superconducting transition temperature. This
casts doubt on whether this material is a simple BCS
superconductor, an indication that electron-electron interaction
may be playing a significant role in this material's
superconducting mechanism.

\section{Acknowledgements}
The authors would like to thank A.~P.~Jephcoat and H. Olijnyk from
the Earth Science Department, Oxford University, UK and
A.~F.~Goncharov, V.~V.~Struzhkin, Ho-kwang~Mao and R.~J.~Hemley
from the Geophysical Laboratory, Carnegie Institute, Washington,
USA for their stimulating collaboration that resulted in the
measurements~\cite{Klehe_JPCM2000,McDonald_JPCM2001} that the
analysis in this paper is based upon. This work is supported by the EPSRC (UK).

\end{sloppypar}


\begin{thebibliography}{10}
\bibitem{endnote1}
In this paper we extract an effective mass from the optical data that is
  strongly influenced by interband electronic processes. We argue that in
  contrast to the intraband processes the interband ones are renormalized by
  quasi-particle interactions, see Subsection~\ref{M*}.
\bibitem{endnote2}
The electronic contributions to the spectra are not accurately reproduced using
  a single frequency independent value for the electronic damping.
\end{thebibliography}

\end{document}